\documentclass[sigconf, 11pt, timestamp=false]{acmart}

\AtBeginDocument{%
  \providecommand\BibTeX{{%
    \normalfont B\kern-0.5em{\scshape i\kern-0.25em b}\kern-0.8em\TeX}}}

\setcopyright{acmcopyright}
\copyrightyear{2023}
\acmYear{2023}

\acmConference[CNB-MAC '23]{The Eighth International Workshop on Computational Network Biology: Modeling, Analysis, and Control}{September 3, 2023}{Houston, TX}
\usepackage{subcaption}
\graphicspath{ {./images/} }

\begin{document}

\title[XVir: Transformer Architecture for Identifying Viral Reads from Cancer Samples]{XVir: A Transformer-Based Architecture for Identifying Viral Reads from Cancer Samples}

\author{Shorya Consul}
\orcid{0000-0001-9137-8989}
\affiliation{%
  \institution{UT Austin}
  \city{Austin, Texas}
  \country{USA}
}
\email{shoryaconsul@utexas.edu}

\author{John Robertson}
\affiliation{%
  \institution{UT Austin}
  \city{Austin, Texas}
  \country{USA}
}
\email{john.robertson@utexas.edu}

\author{Haris Vikalo}
\orcid{0000-0002-7945-4114}
\affiliation{%
  \institution{UT Austin}
  \city{Austin, Texas}
  \country{USA}
}
\email{hvikalo@ece.utexas.edu}

\renewcommand{\shortauthors}{Consul, Robertson and Vikalo}
\vspace{0.5in}

\begin{abstract}
It is estimated that approximately $15$\% of cancers worldwide can be linked to viral infections. The viruses that can cause or increase the risk of cancer include human papillomavirus, hepatitis B and C viruses, Epstein-Barr virus, and human immunodeficiency virus, to name a few. The computational analysis of the massive amounts of tumor DNA data, whose collection is enabled by the recent advancements in sequencing technologies, have allowed studies of the potential association between cancers and viral pathogens. However, the high diversity of oncoviral families makes reliable detection of viral DNA difficult and thus, renders such analysis challenging. In this paper, we introduce XVir, a data pipeline that relies on a transformer-based deep learning architecture to reliably identify viral DNA present in human tumors. In particular, XVir is trained on genomic sequencing reads from viral and human genomes and may be used with tumor sequence information to find evidence of viral DNA in human cancers. Results on semi-experimental data demonstrate that XVir is capable of achieving high detection accuracy, generally outperforming state-of-the-art competing methods while being more compact and less computationally demanding.
\end{abstract}

\ccsdesc[500]{Computer systems organization~Embedded systems}
\ccsdesc[300]{Computer systems organization~Redundancy}
\ccsdesc{Computer systems organization~Robotics}
\ccsdesc[100]{Networks~Network reliability}

\keywords{transformers, deep learning, oncoviral infection, cancer, DNA sequencing, classification}


\maketitle

\section{Introduction}

A number of viruses are known to cause or increase the risk of cancer, including hepatitis B and C viruses (HBV and HBC, respectively), human papillomavirus (HPV), Epstein-Barr virus (EBV), and human immunodeficiency virus (HIV) \cite{liao2006, mclaughlin2008viruses,mui2017,schiller2021,zapatka2020landscape,cantalupo2018viral}. While the link between certain viruses and cancer is widely recognized, the molecular mechanisms of viral carcinogenesis remain only partly understood. For instance, it is known that viruses associated with human tumor encode viral oncoproteins which impact the regulatory cellular processes in the host cells, ultimately promoting formation of tumor. Furthermore, as those viruses integrate into the host cell genome, they may by chance give proliferative advantage to the host cell, which constitutes another tumor promotion mechanism. Nevertheless, such mechanisms are still only partly understood and require closer attention.


The first step towards understanding the role of a viral infection on cancer is to reliably identify the presence of viral genome(s) in tumor cells. This fails when the oncoviral family is highly divergent from known viral sequences. This is prevalent due to the rapid evolution of viral genomes and the incompleteness of genome databases. To overcome this obstacle, many tools for viral DNA detection have been developed for a variety of sequencing data, including RNA, cDNA and/or amino acid sequences with mixed results. ViFi \cite{nguyen2018vifi} used an ensemble of Hidden Markov models (HMM) built from viral reference genomes and subsequently, identifying viral reads that may have evolved from the genomes. VirFinder \cite{ren2017virfinder} used k-mer frequencies to characterize and identify viral sequences from metagenomic data. More recent approaches have looked to utilize deep learning models to improve viral read identification. ViRNAtrap \cite{elbasir2023deep} used a convolutional neural network (CNN) for viral identification from much shorter RNA sequencing reads. DeepVirFinder \cite{ren2020identifying} utilized a shallower CNN (a network with fewer layers) than viRNAtrap to identify viral contigs from DNA sequences. Notably, DeepViFi \cite{rajkumar2022deepvifi} proposed a hybrid pipeline comprising a transformer and random forest for viral read identification -- instead of learning a dense layer for classification, the transformer is followed by a random forest that facilitates classification of the transformer's embeddings. A main drawback to \cite{rajkumar2022deepvifi} is the model's size -- even though the random forest appears to provide major benefit to the accuracy of classification, the transformer embedding seems exceedingly large for the 1-mer encoding proposed by DeepViFi. Similarly, the shallow CNN adopted by DeepVirFinder is also massive as it makes up for the shallow structure by utilizing numerous convolutional filters (on the order of 1000).


\begin{figure}[h]
  \centering  \includegraphics[width=\linewidth]{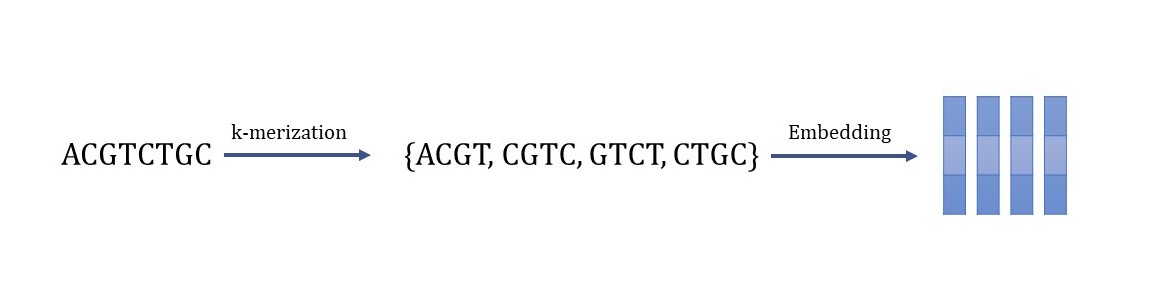}
  \caption{Tokenization of sequencing read into overlapping k-mers. Each k-mer is mapped to an embedding vector $\in \mathbb{R}^{d_m}$. In the given example, $k = 4$.}
  \Description[Worklfow for read embeddings]{Each read is represented as a sequnece of k-mers, and each k-mer is mapped to an embedding.}
  \label{fig:token}
\end{figure}

In this paper, we present a transformer-based deep learning approach where the model is trained on reads obtained by sequencing genomic content of tumor cells and thus does not ignore non-coding regions of DNA but rather expects them to contain information that can assist in viral identification. The remainder of the paper is organized as follows. Section~2 introduces the data processing pipeline, including the transformer architecture and its training procedure. Section~3 reports the experimental settings and the results of benchmarking the proposed method against 
DeepViFi and DeepVirFinder. Finally, Section~4 concludes the paper.

\begin{figure*}
  \centering
  \includegraphics[width=0.95\linewidth]{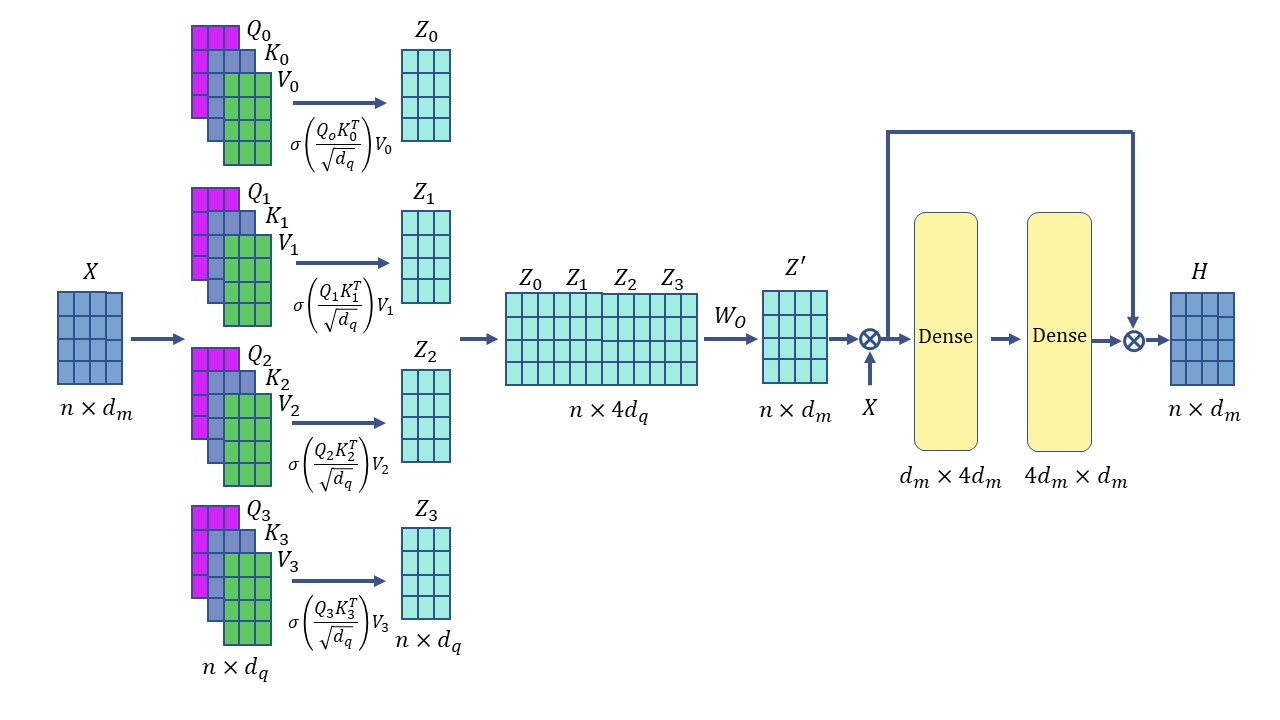}
  \caption{An illustration of the transformer encoder layer. The adder denotes the ``add and layer normalize" operations. $\sigma(.)$ denotes the row-wise softmax operation.}
  \Description[Transformer encoder layer]{Transformer encoder layer}
\label{fig:transformer_encoder}
\end{figure*}

\section{Method}
\subsection{Problem formulation}
High-throughput sequencing platforms are capable of providing massive amounts of short reads. Each sequencing run essentially samples (with replacement) the genome being examined. Therefore, if present, oncoviral infections would manifest themselves via reads of viral origin in the resulting sequencing dataset. Such a set of sequencing reads serves as the input to our pipeline. Viral read identification aims to answers the question: Is a given sequencing read $r_i$ of viral origin?

\subsection{Input preprocessing}
\subsubsection{Read tokenization and embedding} \label{sec:tokenization}
A common method of representing a genomic sequence is by means of the strings of overlapping k-mers, i.e., subsequences of length $k$ \cite{compeau2011bruijn}. Specifically, a read $r_i$ of length $l$ can be represented by enumerating $(l-k+1)$ k-mers starting from the first base in the read. An illustration of this is shown in Fig.~\ref{fig:token}. 
Each k-mer is then mapped to an embedding of size $d_m$. Note that the number of k-mers is exponential in $k$ (specifically, there are $4^k$ k-mers). This implies that one-hot encoding the k-mers requires vectors of size $4^k$. Hence, for larger values of $k$, learning embeddings of size $d_m$ for each k-mer is increasingly more efficient than one-hot encoding.

\subsection{Transformer-based classifier}

This is the main block of the XVir pipeline. It first combines the sequence of k-mer embeddings with position encodings to form a sequence of embeddings $X\in \mathbb{R}^{n\times d_m}$, where each column corresponds to a k-mer infused with positional information and $n=l-k+1$. Positional encodings are required here as the transformer is otherwise agnostic to sequential information; the training of the transformer is invariant to permutations of the input tokens. We elect to use the sinusoidal positional encoding proposed in \cite{vaswani2017attention}:
\begin{align}
    PE_{2j}(pos) &= \sin\left(\frac{pos}{10000^{2j/d_m}}\right) \label{eq:sin}\\
    PE_{2j+1}(pos) &= \cos\left(\frac{pos}{10000^{2j/d_m}}\right) \label{eq:cos}
\end{align}

The positional encoding is added to the read embedding and the result is layer-normalized to obtain $X$. Note that these positional encodings are \textit{not learned}; this helps reduce XVir's model complexity, compared to other transformer-based architectures such as that of DeepViFi without sacrificing performance.

\begin{figure*}
  \centering
  \includegraphics[width=0.8\linewidth]{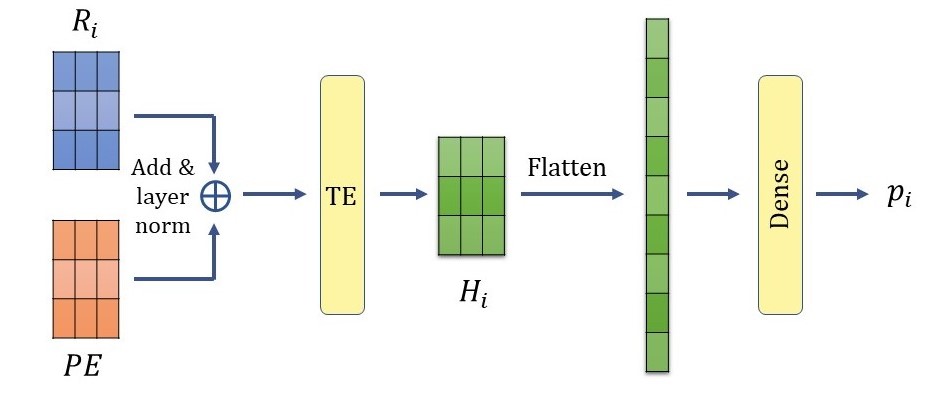}
  \caption{Architecture of classifier. The input (read) is fed into the classifier as a sequence of embeddings and outputs a probability of the read being of viral origin.}
  \Description[Network architecture for XVir]{Transformer-based classifier to classify reads}
  \label{fig:arch}
\end{figure*}

\subsubsection{The transformer encoder layer}
Beginning with a sequence of embeddings, $X$, the transformer encoder employs multi-headed self-attention \cite{vaswani2017attention} to learn read signatures. ``Self-attention" learns how each k-mer is related to the other k-mers in the sequence; we expect that reads of differing origins would exhibit different self-attention patterns, which we shall hereafter refer to as \textit{read signatures}. 

As shown in Fig.~\ref{fig:transformer_encoder}, the transformer encoder quantifies self-attention through the use of query ($Q_i$), key ($K_i$) and value ($V_i$) matrices. These matrices are combined to form self-attention matrices ($Z_i$). Each attention head comprises a different set of these matrices. The self-attention matrices are concatenated and condensed to form the output of the multi-headed self-attention $Z'$. At this point, the input $X$ is added to $Z'$ and layer-normalized before being passed to a feed-forward network. The final output of the encoder is found by adding the input and output of the feed-forward network and layer-normalizing the result. Multiple attention heads improve the likelihood that meaningful read signatures are learned. For instance, an attention head may concentrate most of its weight on a single position in the read, but the other attention heads would likely have better spreads across the read. 

For an input $X\in \mathbb{R}^{n\times d_m}$ where column $X^{j}$ corresponds to the embedding of the $j^{th}$ k-mer in the read, the operation of the transformer encoder with $h$ attention heads can be formalized as
\begin{align}
    Q_i &= XW_i^Q \\
    K_i &= XW_i^K \\
    V_i &= XW_i^V \\
    Z_i &= softmax\left(\frac{Q_iK_i^T}{\sqrt{d_q}}\right) \\
    Z' &= concatenate\left(Z_0, Z_1, Z_2, Z_3\right)W_O \\  
    D_0 &= LayerNorm(X + Z') \\
    D_1 &= ReLU\left(W_1^{dense}\cdot D_0 + B_1^{dense}\right) \\
    D_2 &= ReLU\left(W_2^{dense}\cdot D_1 + B_2^{dense}\right) \\
    H &= LayerNorm(D_0 + D_2)
\end{align}
where the weight matrices, $\left\{W_i^Q, W_i^K, W_i^V\right\}\in \mathbb{R}^{d_m\times d_q}$,  $\left\{Q_i, K_i, V_i\right\}\in \mathbb{R}^{n \times d_q}$, $Z_i\in \mathbb{R}^{n \times d_m}$. XVir employs $h=4$ attention heads, so $d_q = d_m/h = d_m/4$. The output of the multi-headed self-attention $Z'$, as well as $D_0$, $D_2$ and $H$, have the same dimensions as $X$, while $D_1\in\mathbb{R}^{n \times 4d_m}$. The feed-forward network consists of weights $\{W_1^{dense}, W_2^{dense}\}$, biases $\{B_1^{dense}, B_2^{dense}\}$, and uses the ReLU activation function \cite{nair2010rectified}.

\subsubsection{The classifier}
The classifier in XVir utilizes a single transformer encoder layer (see Fig.~\ref{fig:arch}). The output of the classifier, $p_i$, is obtained by taking the sigmoid of a weighted sum of the read signature $H_i$. Denoting the transformer encoder as $TE(.)$, we can formalize the operations of XVir as
\begin{align}
    X_i &= LayerNorm(R_i + PE) \label{eq:embed} \\
    H_i &= TE(X_i) \label{eq:read_sig} \\
    p_i &= \sigma\left(W_C^{dense}\cdot Flatten(H_i) + B_C^{dense}\right) \label{eq:class_dense}
\end{align}
where $R_i$ is an $n\times d_m$ matrix with the $j^{th}$ row corresponding to the embedding of the $j^{th}$ k-mer in the read $i$. $W_C^{dense}$ and $B_C^{dense}$ denote the weights and bias of the final dense layer in XVir. Note that the operation of the dense layer in \eqref{eq:class_dense} can also be expressed as $\Sigma_{s,t} w_{st}^CH_{i,st} + b_{s,t}^C$, i.e., a weighted sum of all the elements of $H_i$ (plus a bias).

\subsection{Network optimization}
Observe that at its core, the problem of viral read identification from a human sample is a binary classification problem. We label reads of viral origin viral as class `1' and other reads as class `0', XVir employs the binary cross-entropy loss to train the network. To that end, we employ the binary cross-entropy loss. Denoting the class of read $i$ as $y_i \in \{0,1\}$, and the predicted probability of the read being of viral origin $p_i$, the binary cross entropy loss is
\begin{equation}
    L = -\sum_i  \left[y_i \log(p_i) + (1-y_i)\log(1-p_i)\right]
\end{equation}

\paragraph{Regularization} We use dropout \cite{hinton2012improving} with a probability of 0.1 when training the network.

\subsection{Hyperparameter tuning}
XVir is trained with the Adam optimizer \cite{kingma2014adam} with a learning rate of $0.001$ and a weight decay of $10^{-6}$. We varied the length of k-mers used from $3$ to $7$, and tested XVir with latent dimension, $d_m \in \{64, 128, 256\}$ and observed the accuracy on the validation data. Based on this, we elected to use $k=6, d_m=128$ as larger values of $k$ and $d_m$ led to only marginal gains in performance on the validation data. XVir was trained for 25 epochs as validation accuracy consistently plateaued after this point.

\paragraph{Computational platform} All the included experiments were performed on a server equipped with 96 1.50 MHz EPYC 7642 processors, 503GB of RAM and AMD Vega 20 GPUs. All models were trained on a single GPU when possible.

\begin{figure}[htb]
    \centering
    \begin{subfigure}[b]{0.45\linewidth}
        \includegraphics[width=\linewidth]{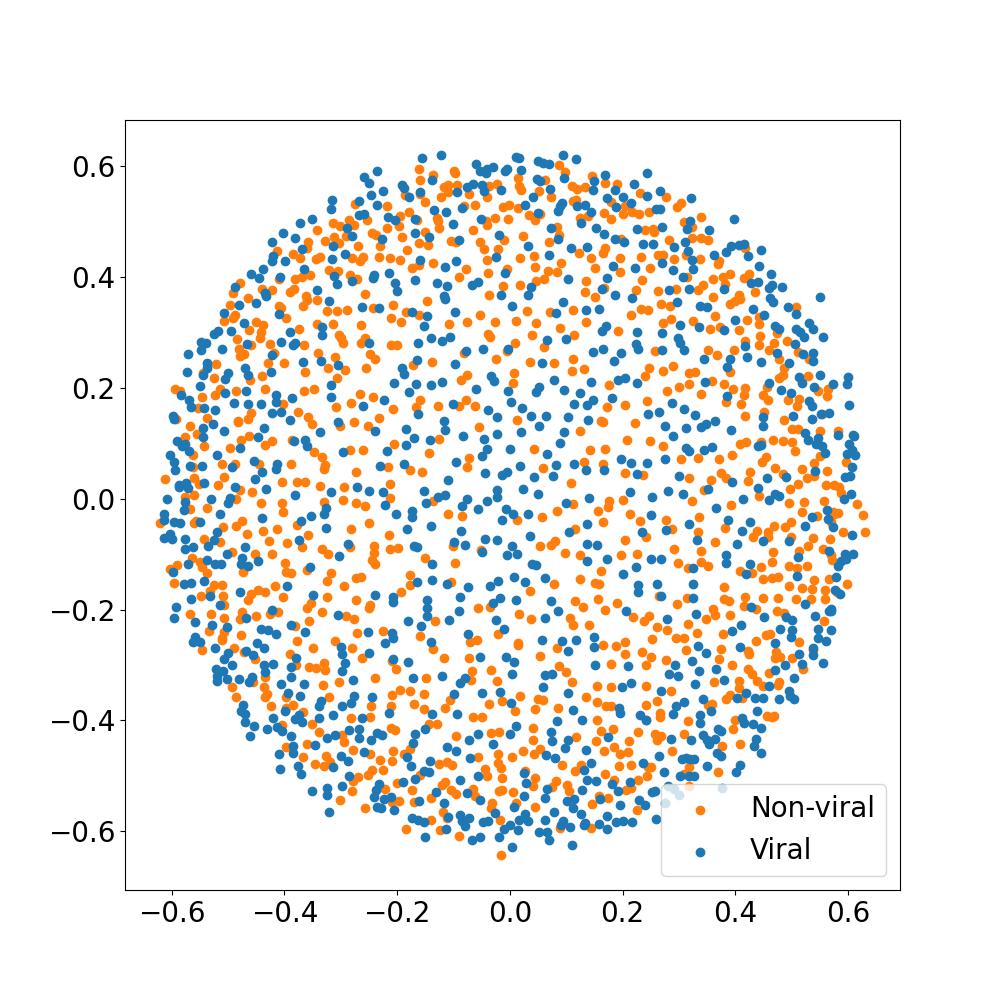}
        \caption{MDS}
        \Description[MDS]{Reads representations obtained via multi-dimensional scaling}
        \label{fig:mds}
    \end{subfigure}
    \begin{subfigure}[b]{0.45\linewidth}
        \includegraphics[width=\linewidth]{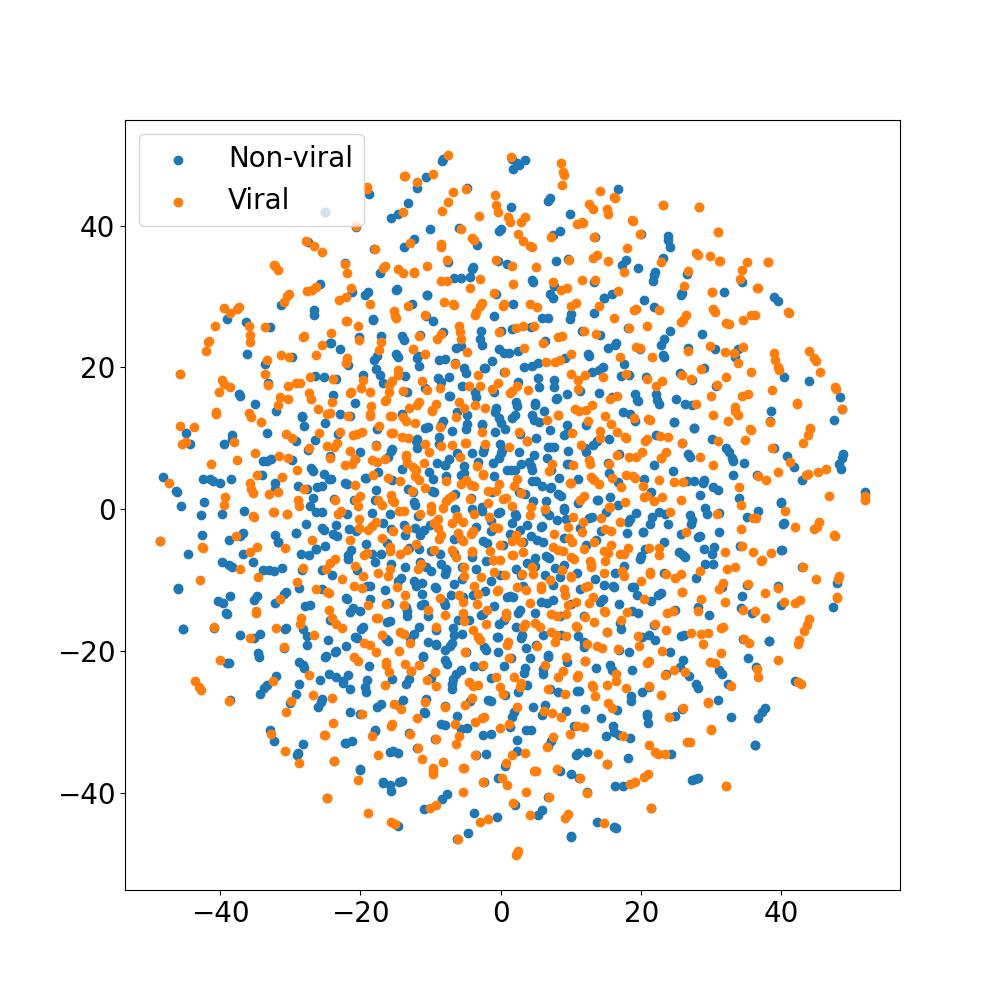}
        \caption{t-SNE}
        \Description[t-SNE]{Reads representations obtained via t-SNE}
    \label{fig:tsne}
    \end{subfigure}
    \caption{Visualization of reads using multidimensional scaling (MDS) and t-SNE. Both visualizations show that viral and non-viral reads are not easily separable.}
    \label{fig:data_vis}
\end{figure}

\section{Results}
\subsection{Experimental settings} \label{sec:exp_setting}
We collected 444 Human Papillomavirus (HPV) genomes from the Papillomavirus Episteme (PaVE) \cite{van2017papillomavirus} as the set of viral reference genomes for our dataset. We also selected all the primary assemblies from the GRCh38.p14 genome assembly \cite{cole2008finishing} as the human reference genomes. ART \cite{huang2012art}is then used to generated 150bp HiSeq 2500 reads to generate the same number of reads from the viral and human genomes. This yields 299,691 reads from the HPV genomes and 278,591 reads from the human genome. These reads were then combined to form our dataset. We partition this dataset into the training, validation and test datasets in the ratio of 8:1:1. While XVir's model parameters are learned using the training data, hyperparameter tuning and model selection are accomplished through the use of validation data. The test set is used to compare the performance of XVir to pre-existing algorithms for oncoviral read identification. The difficulty of this task can be visualized through the use of multidimensional scaling \cite{borg2005modern} and t-SNE \cite{van2008visualizing} to obtain low-dimensional representations of the data. Using Hamming distance as the pairwise distance metric for read pairs, two-dimensional representations of our read data are shown in Fig.~\ref{fig:data_vis}. Evidently, the viral reads are not easily separable from their non-viral counterparts. 

\subsection{Comparison with benchmarks}
We compare the performance of XVir against state-of-the-art methods including DeepViFi \cite{rajkumar2022deepvifi} and DeepVirFinder \cite{ren2020identifying} on the aforementioned test dataset. Since DeepVirFinder was trained on contigs, we retrained DeepVirFinder on our training data with the validation data used for early stopping. XVir achieved an accuracy of 0.967 and an area under the ROC curve (AUROC) of 0.995. This is significantly higher than that attained by DeepViFi (0.767, 0.990) and only marginally lower than that of DeepVirFinder (0.995, 1.0). XVir accomplishes this while utilizing a far smaller model than both DeepViFi and DeepVirFinder; our model is less than 25\% the size of either of the existing models (see Table~\ref{tab:model_complexity}). This, in turn, is reflected in the significantly faster training time for XVir; it is 40\% faster than DeepViFi and over $8\times$ faster than DeepVirFinder. This is especially notable as larger models are more likely to overfit to the training data distribution and more likely to yield errors on unseen data. Such a scenario is likely in oncoviral read identification when contaminant reads, such as fungal or bacterial reads, are commonly present in the tumor sequencing data. Moreover, the use of transformers in XVir is preferable to the convolutional neural network (CNN) utilized by DeepVirFinder as (a) it enables better parallelization, thereby improving GPU utilization for training speedup and, (b) it is better equipped to capture long-range dependencies in sequencing reads as the attention mechanism has a larger `receptive field' than the convolution operation; each attention operation captures information from all positions across the read, while a convolution operation captures information only from the read positions within its filter length.

\begin{figure*}[tb]
    \centering
    \begin{subfigure}{0.3\linewidth}
        \includegraphics[width=\linewidth]{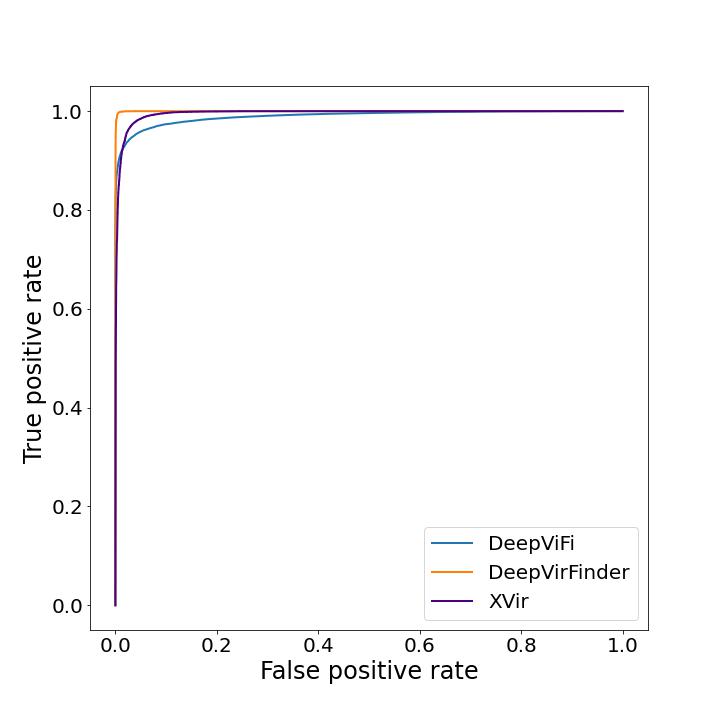}
        \caption{ROC}
        \label{fig:res}
    \end{subfigure}
    \begin{subfigure}{0.3\linewidth}
        \includegraphics[width=\linewidth]{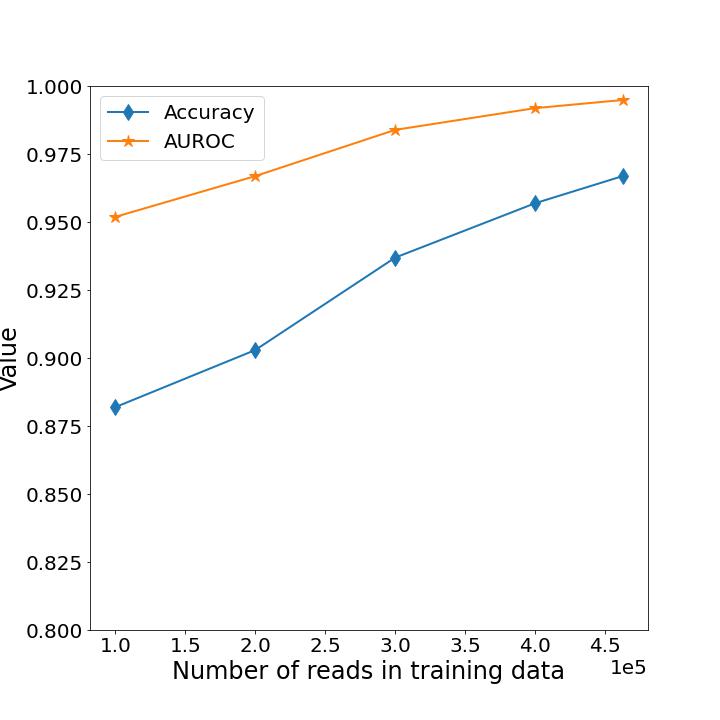}
        \caption{Varying amount of training data}
        \label{fig:data_prevalence}
    \end{subfigure}
    \begin{subfigure}{0.3\linewidth}
        \includegraphics[width=\linewidth]{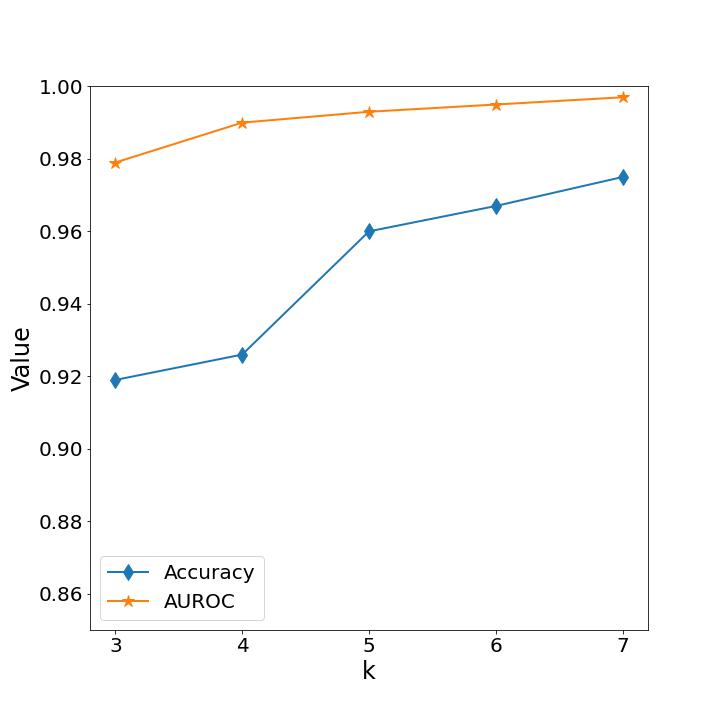}
        \caption{Effect of length of k-mer}
        \label{fig:kmer-perf}
    \end{subfigure}
    \caption{Analysis of the performance of XVir. (A) Receiver operating characteristic of XVir, DeepViFi and DeepVirFinder. (b) Performance of XVir trained on varying number of reads. (c) Effect of length of k-mer used for tokenization on performance of XVir.}
\end{figure*}

\begin{table}
  \caption{Comparison of the model complexity of XVir, DeepViFi and DeepVirFinder. The training times were collected from runs on only the CPU. *XVir required 245s per epoch of training on a single GPU. $^\dag$DeepViFi numbers represent training only the transformer; the random forest is trained separately.}
  \label{tab:model_complexity}
  \begin{tabular}{lcc}
    \toprule
    Method & Parameters & Training time/epoch (in s)\\
    \midrule
    XVir & 741,377 & 665* \\
    DeepViFi & 3,207,942 & $5393^\dag$\\
    DeepVirFinder & 3,129,014& 1037\\
  \bottomrule
\end{tabular}
\end{table}

\subsection{Robustness to data prevalence}
We turn our attention to the amount of data that XVir requires to learn to accurately identify viral reads from an admixture of oncoviral and human reads. We use the validation and test data outlined in Section~\ref{sec:exp_setting} but randomly subsample a fixed number of reads from the training data to form our new training data. Specifically, we sample 100k, 200k, 300k and 400k reads from the training data. All the other hyperparameters of XVir are kept unchanged. As can be seen from Fig.~\ref{fig:data_prevalence}, the performance of XVir on the test data increases monotonically with number of reads in the training data; there is a notable increase in accuracy with more training data. On the other hand, XVir attains an AUROC exceeding 0.95 with just 100,000 reads. This suggests that XVir quickly learns to distinguish viral reads from non-viral ones but requires more reads to calibrate its output class probability, i.e., outputting a class probability $>0.5$ when the read is of viral origin and $\leq 0.5$ otherwise.

\subsection{Effect of k-mer length}
Recall from Section~\ref{sec:tokenization} that a read of length $l$ is tokenized into $n=l-k+1$ k-mers, We analyze the effect of using a richer set of tokens, i.e., a larger $k$, on the performance of XVir on test data in Fig.~\ref{fig:kmer-perf}. Clearly, both the accuracy and AUROC of XVir increase monotonically with $k$, albeit at a diminishing rate. However, the model complexity scales exponentially with $k$ as the number of possible k-mers scales as $4^k$. In particular, the portion of XVir used to learn these k-mer embeddings has $N_e = \mathcal{O}(4^k d_m)$ parameters. For instance, for $k=6$ and $d_m=128$, $N_e = 524,288$; this is over 70\% of the model complexity. Such a trade-off between model complexity and performance is endemic to the paradigm of machine learning \cite{esposito2000approximation}. It is notable that despite the model complexity ballooning with higher values of $k$, the training time for XVir stays almost constant.

\section{Conclusion}
In this paper, we presented a novel algorithm for oncoviral read detection from human host genome sequencing data. Xvir leverages transformers to learn read signatures, which are then used to infer the origin of the read. This is accomplished by first tokenizing each read into a sequence of overlapping k-mers, each of which is embedded in a $d_m$-dimensional space. Benchmarking on semi-experimental data illustrates the superior performance of XVir over a recent transformer-based approach, DeepViFi, and comparable performance to DeepVirFinder, a CNN-based approach towards identifying viral contigs. Moreover, XVir achieves this performance despite having <25\% the number of parameters as both DeepViFi and DeepVirFinder; this also enables faster training of XVir. Due to the larger receptive field of transformers over CNNs, it is likely that XVir may be able to deliver even better classification performance over longer reads/contigs, as well as utilize reads from newer long-read sequencing platforms, such as those from Pacific Biosciences and Oxford Nanopore Technologies \cite{antipov2022viralflye}.

\paragraph{Availability:} The code for XVir is publicly available at \url{https://github.com/shoryaconsul/XVir}. 
\begin{acks}
This work was funded in part by the NSF grant CCF-2109983.
\end{acks}

\bibliographystyle{ACM-Reference-Format}
\bibliography{refs}


\begin{thebibliography}{23}


\ifx \showCODEN    \undefined \def \showCODEN     #1{\unskip}     \fi
\ifx \showDOI      \undefined \def \showDOI       #1{#1}\fi
\ifx \showISBNx    \undefined \def \showISBNx     #1{\unskip}     \fi
\ifx \showISBNxiii \undefined \def \showISBNxiii  #1{\unskip}     \fi
\ifx \showISSN     \undefined \def \showISSN      #1{\unskip}     \fi
\ifx \showLCCN     \undefined \def \showLCCN      #1{\unskip}     \fi
\ifx \shownote     \undefined \def \shownote      #1{#1}          \fi
\ifx \showarticletitle \undefined \def \showarticletitle #1{#1}   \fi
\ifx \showURL      \undefined \def \showURL       {\relax}        \fi
\providecommand\bibfield[2]{#2}
\providecommand\bibinfo[2]{#2}
\providecommand\natexlab[1]{#1}
\providecommand\showeprint[2][]{arXiv:#2}

\bibitem[Antipov et~al\mbox{.}(2022)]%
        {antipov2022viralflye}
\bibfield{author}{\bibinfo{person}{Dmitry Antipov}, \bibinfo{person}{Mikhail
  Rayko}, \bibinfo{person}{Mikhail Kolmogorov}, {and} \bibinfo{person}{Pavel~A
  Pevzner}.} \bibinfo{year}{2022}\natexlab{}.
\newblock \showarticletitle{viralFlye: assembling viruses and identifying their
  hosts from long-read metagenomics data}.
\newblock \bibinfo{journal}{\emph{Genome Biology}} \bibinfo{volume}{23},
  \bibinfo{number}{1} (\bibinfo{year}{2022}), \bibinfo{pages}{1--21}.
\newblock


\bibitem[Borg and Groenen(2005)]%
        {borg2005modern}
\bibfield{author}{\bibinfo{person}{Ingwer Borg} {and}
  \bibinfo{person}{Patrick~JF Groenen}.} \bibinfo{year}{2005}\natexlab{}.
\newblock \bibinfo{booktitle}{\emph{Modern multidimensional scaling: Theory and
  applications}}.
\newblock \bibinfo{publisher}{Springer Science \& Business Media}.
\newblock


\bibitem[Cantalupo et~al\mbox{.}(2018)]%
        {cantalupo2018viral}
\bibfield{author}{\bibinfo{person}{Paul~G Cantalupo}, \bibinfo{person}{Joshua~P
  Katz}, {and} \bibinfo{person}{James~M Pipas}.}
  \bibinfo{year}{2018}\natexlab{}.
\newblock \showarticletitle{Viral sequences in human cancer}.
\newblock \bibinfo{journal}{\emph{Virology}}  \bibinfo{volume}{513}
  (\bibinfo{year}{2018}), \bibinfo{pages}{208--216}.
\newblock


\bibitem[Cole et~al\mbox{.}(2008)]%
        {cole2008finishing}
\bibfield{author}{\bibinfo{person}{Charlotte~G Cole}, \bibinfo{person}{Owen~T
  McCann}, \bibinfo{person}{John~E Collins}, \bibinfo{person}{Karen Oliver},
  \bibinfo{person}{David Willey}, \bibinfo{person}{Susan~M Gribble},
  \bibinfo{person}{Fengtang Yang}, \bibinfo{person}{Karen McLaren},
  \bibinfo{person}{Jane Rogers}, \bibinfo{person}{Zemin Ning}, {et~al\mbox{.}}}
  \bibinfo{year}{2008}\natexlab{}.
\newblock \showarticletitle{Finishing the finished human chromosome 22
  sequence}.
\newblock \bibinfo{journal}{\emph{Genome Biology}} \bibinfo{volume}{9},
  \bibinfo{number}{5} (\bibinfo{year}{2008}), \bibinfo{pages}{1--11}.
\newblock


\bibitem[Compeau et~al\mbox{.}(2011)]%
        {compeau2011bruijn}
\bibfield{author}{\bibinfo{person}{Phillip~EC Compeau},
  \bibinfo{person}{Pavel~A Pevzner}, {and} \bibinfo{person}{Glenn Tesler}.}
  \bibinfo{year}{2011}\natexlab{}.
\newblock \showarticletitle{Why are de Bruijn graphs useful for genome
  assembly?}
\newblock \bibinfo{journal}{\emph{Nature biotechnology}} \bibinfo{volume}{29},
  \bibinfo{number}{11} (\bibinfo{year}{2011}), \bibinfo{pages}{987}.
\newblock


\bibitem[Elbasir et~al\mbox{.}(2023)]%
        {elbasir2023deep}
\bibfield{author}{\bibinfo{person}{Abdurrahman Elbasir}, \bibinfo{person}{Ying
  Ye}, \bibinfo{person}{Daniel~E Sch{\"a}ffer}, \bibinfo{person}{Xue Hao},
  \bibinfo{person}{Jayamanna Wickramasinghe}, \bibinfo{person}{Konstantinos
  Tsingas}, \bibinfo{person}{Paul~M Lieberman}, \bibinfo{person}{Qi Long},
  \bibinfo{person}{Quaid Morris}, \bibinfo{person}{Rugang Zhang},
  {et~al\mbox{.}}} \bibinfo{year}{2023}\natexlab{}.
\newblock \showarticletitle{A deep learning approach reveals unexplored
  landscape of viral expression in cancer}.
\newblock \bibinfo{journal}{\emph{Nature communications}} \bibinfo{volume}{14},
  \bibinfo{number}{1} (\bibinfo{year}{2023}), \bibinfo{pages}{785}.
\newblock


\bibitem[Esposito et~al\mbox{.}(2000)]%
        {esposito2000approximation}
\bibfield{author}{\bibinfo{person}{Anna Esposito}, \bibinfo{person}{Maria
  Marinaro}, \bibinfo{person}{Domenico Oricchio}, {and} \bibinfo{person}{Silvia
  Scarpetta}.} \bibinfo{year}{2000}\natexlab{}.
\newblock \showarticletitle{Approximation of continuous and discontinuous
  mappings by a growing neural RBF-based algorithm}.
\newblock \bibinfo{journal}{\emph{Neural Networks}} \bibinfo{volume}{13},
  \bibinfo{number}{6} (\bibinfo{year}{2000}), \bibinfo{pages}{651--665}.
\newblock


\bibitem[Hinton et~al\mbox{.}(2012)]%
        {hinton2012improving}
\bibfield{author}{\bibinfo{person}{Geoffrey~E Hinton}, \bibinfo{person}{Nitish
  Srivastava}, \bibinfo{person}{Alex Krizhevsky}, \bibinfo{person}{Ilya
  Sutskever}, {and} \bibinfo{person}{Ruslan~R Salakhutdinov}.}
  \bibinfo{year}{2012}\natexlab{}.
\newblock \showarticletitle{Improving neural networks by preventing
  co-adaptation of feature detectors}.
\newblock \bibinfo{journal}{\emph{arXiv preprint arXiv:1207.0580}}
  (\bibinfo{year}{2012}).
\newblock


\bibitem[Huang et~al\mbox{.}(2012)]%
        {huang2012art}
\bibfield{author}{\bibinfo{person}{Weichun Huang}, \bibinfo{person}{Leping Li},
  \bibinfo{person}{Jason~R Myers}, {and} \bibinfo{person}{Gabor~T Marth}.}
  \bibinfo{year}{2012}\natexlab{}.
\newblock \showarticletitle{ART: a next-generation sequencing read simulator}.
\newblock \bibinfo{journal}{\emph{Bioinformatics}} \bibinfo{volume}{28},
  \bibinfo{number}{4} (\bibinfo{year}{2012}), \bibinfo{pages}{593--594}.
\newblock


\bibitem[Kingma and Ba(2014)]%
        {kingma2014adam}
\bibfield{author}{\bibinfo{person}{Diederik~P Kingma} {and}
  \bibinfo{person}{Jimmy Ba}.} \bibinfo{year}{2014}\natexlab{}.
\newblock \showarticletitle{Adam: A method for stochastic optimization}.
\newblock \bibinfo{journal}{\emph{arXiv preprint arXiv:1412.6980}}
  (\bibinfo{year}{2014}).
\newblock


\bibitem[Liao(2006)]%
        {liao2006}
\bibfield{author}{\bibinfo{person}{John~B Liao}.}
  \bibinfo{year}{2006}\natexlab{}.
\newblock \showarticletitle{Viruses and human cancer.}
\newblock \bibinfo{journal}{\emph{Yale J Biol Med}} \bibinfo{volume}{79},
  \bibinfo{number}{3-4} (\bibinfo{date}{Dec} \bibinfo{year}{2006}),
  \bibinfo{pages}{115--122}.
\newblock


\bibitem[McLaughlin-Drubin and Munger(2008)]%
        {mclaughlin2008viruses}
\bibfield{author}{\bibinfo{person}{Margaret~E McLaughlin-Drubin} {and}
  \bibinfo{person}{Karl Munger}.} \bibinfo{year}{2008}\natexlab{}.
\newblock \showarticletitle{Viruses associated with human cancer}.
\newblock \bibinfo{journal}{\emph{Biochimica et Biophysica Acta (BBA)-Molecular
  Basis of Disease}} \bibinfo{volume}{1782}, \bibinfo{number}{3}
  (\bibinfo{year}{2008}), \bibinfo{pages}{127--150}.
\newblock


\bibitem[Mui et~al\mbox{.}(2017)]%
        {mui2017}
\bibfield{author}{\bibinfo{person}{Uyen~Ngoc Mui},
  \bibinfo{person}{Christopher~T Haley}, {and} \bibinfo{person}{Stephen~K
  Tyring}.} \bibinfo{year}{2017}\natexlab{}.
\newblock \showarticletitle{Viral Oncology: Molecular Biology and
  Pathogenesis.}
\newblock \bibinfo{journal}{\emph{J Clin Med}} \bibinfo{volume}{6},
  \bibinfo{number}{12} (\bibinfo{date}{Nov} \bibinfo{year}{2017}).
\newblock


\bibitem[Nair and Hinton(2010)]%
        {nair2010rectified}
\bibfield{author}{\bibinfo{person}{Vinod Nair} {and}
  \bibinfo{person}{Geoffrey~E Hinton}.} \bibinfo{year}{2010}\natexlab{}.
\newblock \showarticletitle{Rectified linear units improve restricted boltzmann
  machines}. In \bibinfo{booktitle}{\emph{Proceedings of the 27th international
  conference on machine learning (ICML-10)}}. \bibinfo{pages}{807--814}.
\newblock


\bibitem[Nguyen et~al\mbox{.}(2018)]%
        {nguyen2018vifi}
\bibfield{author}{\bibinfo{person}{Nam-Phuong~D Nguyen}, \bibinfo{person}{Viraj
  Deshpande}, \bibinfo{person}{Jens Luebeck}, \bibinfo{person}{Paul~S Mischel},
  {and} \bibinfo{person}{Vineet Bafna}.} \bibinfo{year}{2018}\natexlab{}.
\newblock \showarticletitle{ViFi: accurate detection of viral integration and
  mRNA fusion reveals indiscriminate and unregulated transcription in proximal
  genomic regions in cervical cancer}.
\newblock \bibinfo{journal}{\emph{Nucleic acids research}}
  \bibinfo{volume}{46}, \bibinfo{number}{7} (\bibinfo{year}{2018}),
  \bibinfo{pages}{3309--3325}.
\newblock


\bibitem[Rajkumar et~al\mbox{.}(2022)]%
        {rajkumar2022deepvifi}
\bibfield{author}{\bibinfo{person}{Utkrisht Rajkumar}, \bibinfo{person}{Sara
  Javadzadeh}, \bibinfo{person}{Mihir Bafna}, \bibinfo{person}{Dongxia Wu},
  \bibinfo{person}{Rose Yu}, \bibinfo{person}{Jingbo Shang}, {and}
  \bibinfo{person}{Vineet Bafna}.} \bibinfo{year}{2022}\natexlab{}.
\newblock \showarticletitle{DeepViFi: detecting oncoviral infections in cancer
  genomes using transformers}. In \bibinfo{booktitle}{\emph{Proceedings of the
  13th ACM International Conference on Bioinformatics, Computational Biology
  and Health Informatics}}. \bibinfo{pages}{1--8}.
\newblock


\bibitem[Ren et~al\mbox{.}(2017)]%
        {ren2017virfinder}
\bibfield{author}{\bibinfo{person}{Jie Ren}, \bibinfo{person}{Nathan~A
  Ahlgren}, \bibinfo{person}{Yang~Young Lu}, \bibinfo{person}{Jed~A Fuhrman},
  {and} \bibinfo{person}{Fengzhu Sun}.} \bibinfo{year}{2017}\natexlab{}.
\newblock \showarticletitle{VirFinder: a novel k-mer based tool for identifying
  viral sequences from assembled metagenomic data}.
\newblock \bibinfo{journal}{\emph{Microbiome}} \bibinfo{volume}{5},
  \bibinfo{number}{1} (\bibinfo{year}{2017}), \bibinfo{pages}{1--20}.
\newblock


\bibitem[Ren et~al\mbox{.}(2020)]%
        {ren2020identifying}
\bibfield{author}{\bibinfo{person}{Jie Ren}, \bibinfo{person}{Kai Song},
  \bibinfo{person}{Chao Deng}, \bibinfo{person}{Nathan~A Ahlgren},
  \bibinfo{person}{Jed~A Fuhrman}, \bibinfo{person}{Yi Li},
  \bibinfo{person}{Xiaohui Xie}, \bibinfo{person}{Ryan Poplin}, {and}
  \bibinfo{person}{Fengzhu Sun}.} \bibinfo{year}{2020}\natexlab{}.
\newblock \showarticletitle{Identifying viruses from metagenomic data using
  deep learning}.
\newblock \bibinfo{journal}{\emph{Quantitative Biology}}  \bibinfo{volume}{8}
  (\bibinfo{year}{2020}), \bibinfo{pages}{64--77}.
\newblock


\bibitem[Schiller and Lowy(2021)]%
        {schiller2021}
\bibfield{author}{\bibinfo{person}{John~T Schiller} {and}
  \bibinfo{person}{Douglas~R Lowy}.} \bibinfo{year}{2021}\natexlab{}.
\newblock \showarticletitle{An Introduction to Virus Infections and Human
  Cancer.}
\newblock \bibinfo{journal}{\emph{Recent Results Cancer Res}}
  \bibinfo{volume}{217} (\bibinfo{year}{2021}), \bibinfo{pages}{1--11}.
\newblock


\bibitem[Van~der Maaten and Hinton(2008)]%
        {van2008visualizing}
\bibfield{author}{\bibinfo{person}{Laurens Van~der Maaten} {and}
  \bibinfo{person}{Geoffrey Hinton}.} \bibinfo{year}{2008}\natexlab{}.
\newblock \showarticletitle{Visualizing data using t-SNE.}
\newblock \bibinfo{journal}{\emph{Journal of machine learning research}}
  \bibinfo{volume}{9}, \bibinfo{number}{11} (\bibinfo{year}{2008}).
\newblock


\bibitem[Van~Doorslaer et~al\mbox{.}(2017)]%
        {van2017papillomavirus}
\bibfield{author}{\bibinfo{person}{Koenraad Van~Doorslaer},
  \bibinfo{person}{Zhiwen Li}, \bibinfo{person}{Sandhya Xirasagar},
  \bibinfo{person}{Piet Maes}, \bibinfo{person}{David Kaminsky},
  \bibinfo{person}{David Liou}, \bibinfo{person}{Qiang Sun},
  \bibinfo{person}{Ramandeep Kaur}, \bibinfo{person}{Yentram Huyen}, {and}
  \bibinfo{person}{Alison~A McBride}.} \bibinfo{year}{2017}\natexlab{}.
\newblock \showarticletitle{The Papillomavirus Episteme: a major update to the
  papillomavirus sequence database}.
\newblock \bibinfo{journal}{\emph{Nucleic acids research}}
  \bibinfo{volume}{45}, \bibinfo{number}{D1} (\bibinfo{year}{2017}),
  \bibinfo{pages}{D499--D506}.
\newblock


\bibitem[Vaswani et~al\mbox{.}(2017)]%
        {vaswani2017attention}
\bibfield{author}{\bibinfo{person}{Ashish Vaswani}, \bibinfo{person}{Noam
  Shazeer}, \bibinfo{person}{Niki Parmar}, \bibinfo{person}{Jakob Uszkoreit},
  \bibinfo{person}{Llion Jones}, \bibinfo{person}{Aidan~N Gomez},
  \bibinfo{person}{{\L}ukasz Kaiser}, {and} \bibinfo{person}{Illia
  Polosukhin}.} \bibinfo{year}{2017}\natexlab{}.
\newblock \showarticletitle{Attention is all you need}.
\newblock \bibinfo{journal}{\emph{Advances in neural information processing
  systems}}  \bibinfo{volume}{30} (\bibinfo{year}{2017}).
\newblock


\bibitem[Zapatka et~al\mbox{.}(2020)]%
        {zapatka2020landscape}
\bibfield{author}{\bibinfo{person}{Marc Zapatka}, \bibinfo{person}{Ivan
  Borozan}, \bibinfo{person}{Daniel~S Brewer}, \bibinfo{person}{Murat Iskar},
  \bibinfo{person}{Adam Grundhoff}, \bibinfo{person}{Malik Alawi},
  \bibinfo{person}{Nikita Desai}, \bibinfo{person}{Holger S{\"u}ltmann},
  \bibinfo{person}{Holger Moch}, {et~al\mbox{.}}}
  \bibinfo{year}{2020}\natexlab{}.
\newblock \showarticletitle{The landscape of viral associations in human
  cancers}.
\newblock \bibinfo{journal}{\emph{Nature genetics}} \bibinfo{volume}{52},
  \bibinfo{number}{3} (\bibinfo{year}{2020}), \bibinfo{pages}{320--330}.
\newblock


\end{thebibliography}

\newpage
\appendix

\section{Intuition behind positional encoding}
\begin{figure}
    \centering
    \includegraphics[width=\linewidth]{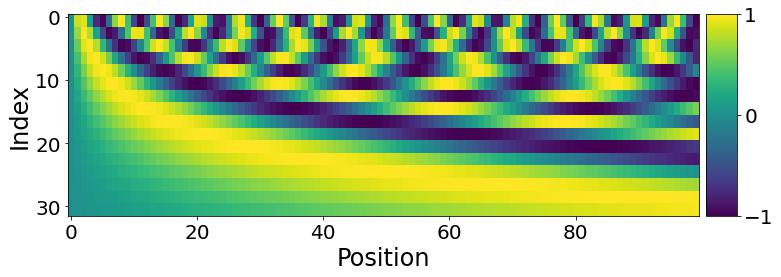}
    \caption{Sinusoidal positional encodings,}
    \label{fig:sin_pos_enc}
\end{figure}

XVir sums the input sequence of embeddings with positional encoding prior to the transformer. This is necessary as the sequence of embeddings do not inherently carry any positional information. This can be formally shown as follows: If the model parameters $M$ minimize the cross-entropy loss for inputs $\mathbf{R}=\{R_i\}$ and $P(.)$ is a permutation over the tokens in each sequence (rows of $R_i$), then $P(M)$ would minimize the cross-entropy loss for input $\tilde{P}(\mathbf{R})$, where $\tilde{P}(M)$ is the permutation over the model parameters corresponding to the permutation $P(.)$ of the input tokens. This is clearly undesirable. \cite{vaswani2017attention} proposed the use of the sinusoidal positional encodings, as specified in \eqref{eq:sin} and \eqref{eq:cos}. Such an encoding can be thought of as a continuous extension of bitstream. As one counts up, the least significant bit in a bitstream flips with the highest frequency, with the flipping frequency decreasing for more significant bits. Fig.~\ref{fig:sin_pos_enc} shows that sinusoidal positional encodings exhibit a smaller pattern - the encoding at the first index has the highest frequency, and this frequency decreases at higher indices in the encoding. The use of sinusoidal positional encodings enable extrapolation to longer sequences, while obviating the need to learn embedding weights for the positional encodings. Instead, these encodings can be pre-computed and used in each epoch.

\end{document}